\PassOptionsToPackage{table,xcdraw}{xcolor}

\documentclass[10pt,conference] {IEEEtran}
\usepackage{tikz}
\usepackage{placeins}
\usetikzlibrary{chains,positioning,shadows}
\usetikzlibrary{shapes.geometric, arrows.meta, positioning}

\tikzstyle{decision} = [diamond, draw, fill=blue!10, text width=4cm, align=center, inner sep=1pt]
\tikzstyle{outcome} = [rectangle, draw, fill=green!10, text width=4cm, align=center, rounded corners, minimum height=1cm]
\tikzstyle{arrow} = [->, thick]

\usepackage[numbers]{natbib}
\usepackage[hidelinks]{hyperref}
\usepackage{xcolor}
\usepackage{fancybox}
\usepackage{multirow}
\usepackage{flushend}
\usepackage{booktabs}
\usepackage{tabularx}
\usepackage{comment}
\usepackage{array}
\usepackage[flushleft]{threeparttable}
\usepackage{mdframed}
\graphicspath{{}{images/}{dia/}}
\DeclareGraphicsExtensions{.pdf,.png}

\usepackage{listings}
\usepackage{courier}
\usepackage[normalem]{ulem}
\usepackage{color}
\usepackage{hyphenat}
\usepackage{alltt                                    
	, multirow
	, booktabs
	, listings
	,float
	,verbatim
	,mathtools
	,url
	,amsmath
}
\usepackage{framed,lipsum}
\usepackage[most]{tcolorbox}
\usepackage[table]{xcolor}
\usepackage{graphics} 
\usepackage{graphicx}
\usepackage{pgfplots, pgfplotstable}
\pgfplotsset{compat=1.15}
\usepackage{tikz}
\usetikzlibrary{patterns}

\usepackage{algorithm2e}
\usepackage{algpseudocode}

\newcounter{o}
\usepackage{xspace}

\usepackage[export]{adjustbox}

\usepackage{tikz}
\usepackage{pgf-pie}
\usetikzlibrary{positioning,shadows}

\usepackage{pgfplots}
\usepackage{pgfplotstable}
\usepackage{adjustbox}
\pgfplotsset{compat=1.18}
\usepackage{sansmath}
\usepackage{tikz}
\usetikzlibrary{positioning}

\newif\ifpienumberinlegend
\pgfkeys{/number in legend/.code=
    \expandafter\let\expandafter\ifpienumberinlegend
    \csname if#1\endcsname
    \ifpienumberinlegend

    \def\beforenumber##1\afternumber{}%
    \fi,
    /number in legend/.default=true
}

\definecolor{1c1}{RGB}{188,162,6}
\definecolor{1c2}{RGB}{137,129,80}
\definecolor{1c3}{RGB}{239,167,31}
\definecolor{1c4}{RGB}{88,194,241}
\definecolor{1c5}{RGB}{6,180,188}

\tikzset{mynode/.style={draw=white,solid,circle,fill=green,inner sep=1pt, thick,
text=black}}
\tikzset{arrow line/.style={dashed, line width= 2.5pt, color=#1}}

\usepackage{paralist}

\usepackage{balance}

\usepackage{listings}
\usepackage{xcolor}
\usepackage{caption}
\usepackage{subcaption}  

\definecolor{keywordcolor}{rgb}{0.0,0.2,0.8}
\definecolor{bgcolor}{rgb}{0.98,0.98,0.98}
\definecolor{framecolor}{rgb}{0.85,0.85,0.85}

\lstdefinestyle{javaStyle}{
  language=Java,
  backgroundcolor=\color{bgcolor},
  basicstyle=\ttfamily\tiny,
  keywordstyle=\color{keywordcolor}\bfseries,
  commentstyle=\color{gray}\itshape,
  stringstyle=\color{red!50!brown},
  showstringspaces=false,
  breaklines=true,
  frame=single,
  rulecolor=\color{framecolor},
  framerule=0.3pt,
  xleftmargin=0pt,
  framexleftmargin=0pt,
  framexrightmargin=0pt,
  framextopmargin=4pt,
  framexbottommargin=4pt,
  tabsize=2
}

\definecolor{keycolor}{rgb}{0.5,0.0,0.5}   
\definecolor{boolcolor}{rgb}{0.0,0.2,0.8}  
\definecolor{strcolor}{rgb}{0.4,0.1,0.1}   
\definecolor{nullcolor}{RGB}{105,105,105}  
\definecolor{linkcolor}{RGB}{139,0,0}      

\lstdefinestyle{yamlStyle}{
  backgroundcolor=\color{bgcolor},
  basicstyle=\ttfamily\tiny,
  keywordstyle=\color{keywordcolor}\bfseries,
  commentstyle=\color{gray}\itshape,
  stringstyle=\color{red!50!brown},
  showstringspaces=false,
  breaklines=true,
  frame=single,
  rulecolor=\color{framecolor},
  framerule=0.3pt,
  xleftmargin=0pt,
  framexleftmargin=0pt,
  framexrightmargin=0pt,
  framextopmargin=4pt,
  framexbottommargin=4pt,
  tabsize=2,
  morekeywords={true,false,null},
  keywordstyle=\color{boolcolor}\bfseries,
  morekeywords={
    api,violations,crash,description,location,file,method,
    fix,commit,revision,internal,pattern,report,source,name,url},
  keywordstyle=\color{keycolor}\bfseries,
}

\normalsize

\usepackage{enumitem}

\usepackage{tikz}

\tcbset{valign=center, add to natural height=0.4cm, blanker, borderline west={1mm}{0mm}{gray},  interior engine=spartan, colback=gray!10, add to height=1cm}
\newtcolorbox[]{findingbox}[1][] { #1 }
\newtcolorbox[]{promptbox}[1][] { reset, #1}

\lstdefinestyle{inlinecode}{basicstyle={\ttfamily\scriptsize\bfseries}}

\newcommand{\urls}[1]{{\scriptsize\url{#1}}}
\usepackage{tcolorbox}

\usepackage{soul}
\usepackage{paralist}
\usepackage[outercaption]{sidecap}    

\usepackage{enumitem}
\usepackage[T1]{fontenc}
\usepackage{pifont}

\definecolor{lightgray}{gray}{.92}

\newtcolorbox
{mybox}[2][]{colbacktitle=red!10!white,
colback=blue!10!white,coltitle=black!70!black,
title={#2},fonttitle=\bfseries,#1}
\usepackage{graphicx}

\definecolor{Gray}{gray}{0.9}

\newboolean{showcomments}
\setboolean{showcomments}{true} 

\ifthenelse{\boolean{showcomments}}
{
  \newcommand{\nbc}[3]{
    \colorbox{#3}{\bfseries\sffamily\scriptsize\textcolor{white}{#1}}
    {\textcolor{#3}{\sf\small$\blacktriangleright$\textit{#2}$\blacktriangleleft$}}
  }
}
{
  \newcommand{\nbc}[3]{}
}

\begin{document}

\title{Can We Trust the AI Pair Programmer? Copilot for API Misuse Detection and Correction}


\author{
\IEEEauthorblockN{Saikat Mondal\hspace{3mm} Chanchal K. Roy\hspace{3mm} Hong Wang\hspace{3mm} Juan Arguello\hspace{3mm} Samantha Mathan}
\IEEEauthorblockA{Department of Computer Science, University of Saskatchewan, Canada\\
\ \{saikat.mondal, chanchal.roy, how763, xgr074, stm875\}@usask.ca}
}

\maketitle

\begin{abstract} 

API misuse introduces security vulnerabilities, system failures, and increases maintenance costs,
all of which remain critical challenges in software development. Existing detection approaches rely on static analysis or machine learning-based tools that operate post-development, which delays defect resolution.  
Delayed defect resolution can significantly increase the cost and complexity of maintenance and negatively impact software reliability and user trust. AI-powered code assistants, such as GitHub Copilot, offer the potential for real-time API misuse detection within development environments. This study evaluates GitHub Copilot’s effectiveness in identifying and correcting API misuse using MUBench, which provides a curated benchmark of misuse cases. 
We construct 740 misuse examples, manually and via AI-assisted variants, using correct usage patterns and misuse specifications. These examples and 147 correct usage cases are analyzed using Copilot integrated in Visual Studio Code. Copilot achieved a detection accuracy of 86.2\%, precision of 91.2\%, and recall of 92.4\%. It performed strongly on common misuse types (e.g., missing/call, null\_check) but struggled with compound or context-sensitive cases. Notably, Copilot successfully fixed over 95\% of the misuses it identified.
These findings highlight both the strengths and limitations of AI-driven coding assistants, positioning Copilot as a promising tool for real-time pair programming and detecting and fixing API misuses during software development.

\end{abstract}



\maketitle

\section{Introduction}
\label{sec:introduction}

Application Programming Interface (API) misuse refers to incorrect usages of an API, that is, violations of their usage constraints, such as call order or preconditions \cite{sven2019investigating}. 
API misuse is still a prevalent issue for software development, as it leads to bugs, security vulnerabilities, data loss, and unexpected system crashes due to errors thrown by not following API usage rules \cite{wei2024demystifying}. 
For example, a study by Li et al.~\cite{li2021large} on GitHub repositories analyzed over one million bug-fix commits and found that 57.1\% of them were caused by API misuses. 
Similarly, studies have shown that defects introduced by API misuse in software projects degrade their quality and can cost millions of dollars to a country's economy \cite{Krasner_2022}. Against this backdrop, growing interest is in leveraging AI-assisted programming tools to automatically detect and prevent API misuses, thereby improving software reliability by reducing costly defects.

GitHub Copilot\footnote{https://github.com/features/copilot} is an AI-powered coding assistant that can support context-aware code completion or generation and interactive feedback as developers write code. It analyzes the local context, including comments, function signatures, and surrounding code, to generate real-time recommendations. When integrated into code editors like Visual Studio Code, users can also open its chat window and ask questions directly about any code file in the current repository. 

Prior studies primarily evaluate GitHub Copilot's capabilities in program synthesis \cite{mastropaolo2023robustness, moassessing} and assessing the quality of its generated code \cite{yetistiren2022assessing, imai2022github, nguyen2022empirical, dakhel2023github, wermelinger2023using, moassessing}. Besides these, a few studies have explored its effectiveness in generating test cases \cite{mehmood2023manual, el2024using, alves2024detecting}.
While the usefulness of Copilot is evident in program synthesis, pair programming, and test case generation, its capabilities have been limitedly investigated in other critical areas. In particular, its ability to detect and fix API misuse in real-world code contexts remains underexplored, an important gap given the prevalence and impact of such misuse in software development.

This gap is especially significant given the rising cost of detecting and fixing API misuse later in the software lifecycle, as is often the case with static analysis and machine learning-based tools that detect such issues only after the code is written \cite{9885793, masood2024statictoolsevaluatinglarge, xia2024exploringautomaticcryptographicapi}. Studies have shown that addressing software defects during the testing phase can cost 15 times more and up to 100 times more during the maintenance phase, compared to fixing them during implementation \cite{Dawson2010}. The exponential increase in the cost of late-stage bug fixes underscores the importance of investigating whether tools like Copilot can facilitate the early detection and resolution of critical issues, such as API misuse.

In this study, we investigate GitHub Copilot's effectiveness in detecting and fixing API misuse during software development. 
First, we constructed examples of API misuse using both manual and AI-assisted methods. These examples were derived from the MUBench dataset~\cite{amann2016mubench}, which contains correct API usage examples from real-world Java projects and documents API misuse scenarios in \texttt{misuse.yml} files.
We then examined GitHub Copilot's ability to detect API misuse by integrating it as an extension in Visual Studio Code. We assessed how accurately it identifies misuse patterns when presented with incorrect API usage in realistic coding contexts. Finally, we evaluated Copilot's capability to fix the misused code by generating appropriate corrections for the identified misuse scenarios. In particular, we answered three research questions and thus made three contributions to this study.

\begin{figure}[!t]
\centering
\includegraphics[width=3.45in]{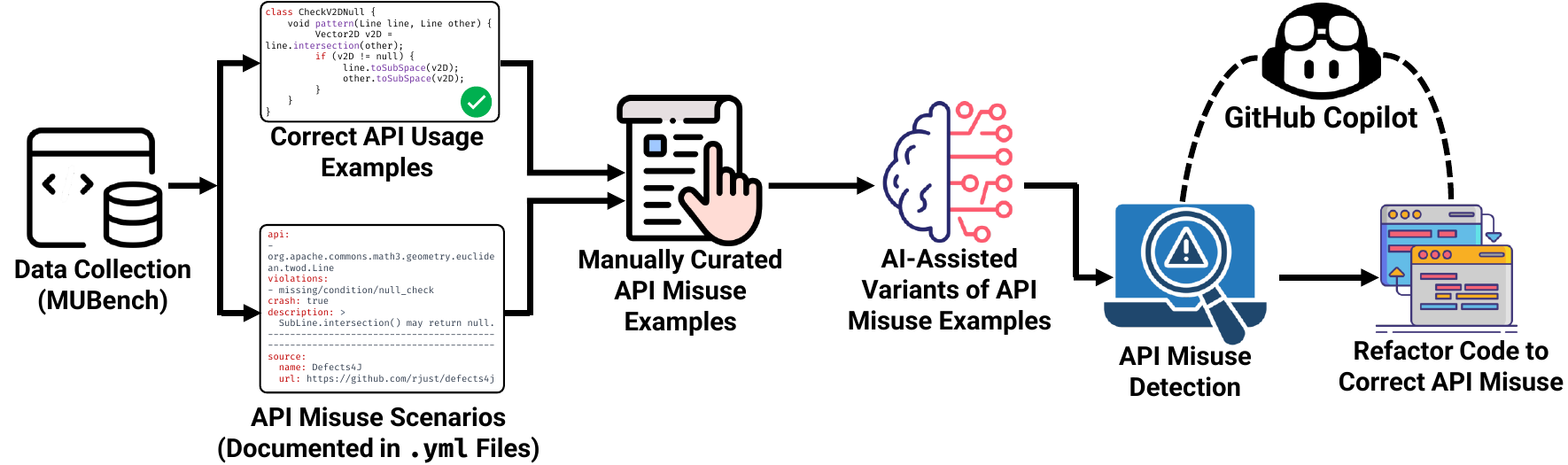} 
\caption{Study Methodology.}
\label{fig:methodology}
\end{figure}

\textbf{RQ1: To what extent can GitHub Copilot accurately identify API misuse patterns while code is being written in real time?}  
This question assesses Copilot’s reliability in detecting API misuses during coding. Evaluating its accuracy is crucial for understanding whether Copilot can serve as an effective early warning tool for developers and reduce reliance on traditional post-hoc static analysis techniques.

\textbf{RQ2: Which types of API misuse are most effectively detected by GitHub Copilot?}
This question explores whether Copilot is more effective at detecting specific categories of API misuse (e.g., missing method calls, incorrect call sequences, or invalid parameter values) than others. Identifying these patterns can guide future improvements and help developers better understand Copilot’s strengths and limitations.

\textbf{RQ3: To what extent can GitHub Copilot refactor code to fix API misuse patterns?}
Beyond detection, this question explores Copilot’s ability to suggest meaningful fixes for API misuse. It focuses on whether the suggested corrections are both syntactically valid and semantically appropriate, potentially reducing developers’ effort in manual debugging and refactoring.

  
  
  

\noindent Our study findings offer empirical insights into GitHub Copilot's effectiveness in detecting and fixing API misuses. By assessing its accuracy, coverage across misuse types, and ability to suggest corrections, we evaluate Copilot's reliability as a real-time assistant and its potential to address gaps left by traditional post-hoc detection tools that detect such issues only after the code is written.

\smallskip
The \textbf{Replication Package} can be found in our online appendix~\cite{replication}.

\begin{figure}[htbp]
  \centering

  \begin{subfigure}{3in}
    \centering
    \begin{lstlisting}[style=javaStyle]
import org.apache.commons.math3.geometry.euclidean.threed.Line;
import org.apache.commons.math3.geometry.euclidean.threed.Vector3D;

class CheckV3DNull {
  void pattern(Line line, Line other) {
    Vector3D v1D = line.intersection(other);
    if (v1D != null) {
      line.toSubSpace(v1D);
      other.toSubSpace(v1D);
    }
  }
}
    \end{lstlisting}
    \vspace{-2mm}
    \caption{Correct usage example.}
    \label{fig:correct-example}
  \end{subfigure}

  \begin{subfigure}{3in}
    \centering
    \begin{lstlisting}[style=yamlStyle]
api:
- org.apache.commons.math3.geometry.euclidean.threed.Line
violations:
- missing/condition/null_check
crash: true
description: >
  SubLine.intersection() may return null.
location:
  file: org/apache/commons/math3/geometry/euclidean/threed/SubLine.java
  method: "intersection(SubLine, boolean)"
fix:
  commit: http://svn.apache.org/viewvc/commons/proper/math/trunk/src/main/java/org/apache/commons/math3/geometry/euclidean/threed/SubLine.java?r1=1488866&r2=1488865&pathrev=1488866&diff_format=h
  description: >
    Check result for null before using.
  revision: d270055e874148a2742604be36ab977eec030fba
internal: true
pattern:
- single object
report: https://issues.apache.org/jira/browse/MATH-988
source:
  name: Defects4J
  url: https://github.com/rjust/defects4j
    \end{lstlisting}
    \vspace{-2mm}
    \caption{Misuse metadata (YAML file content).}
    \label{fig:misuse-yml}
  \end{subfigure}

  \begin{subfigure}{3in}
    \centering
    \begin{lstlisting}[style=javaStyle]
import org.apache.commons.math3.geometry.euclidean.threed.Line;
import org.apache.commons.math3.geometry.euclidean.threed.Vector3D;

class CheckV3DNull {
    void pattern(Line line, Line other) {
        Vector3D v1D = line.intersection(other);
        line.toSubSpace(v1D);
        other.toSubSpace(v1D);
    }
}
    \end{lstlisting}
    \vspace{-2mm}
    \caption{Created misuse example.}
    \label{fig:misuse-example}
  \end{subfigure}
  
  \caption{Example of API misuse created from correct usage and misuse metadata.}
  \label{fig:misuse-example-using-yaml}
  \vspace{-4mm}
\end{figure}

\section{Methodology}
\label{sec:methodology}

Fig. \ref{fig:methodology} shows the schematic diagram of our study methodology. The following sections discuss the different steps of our methodology.

\subsection{Dataset Construction}
\label{subsec:dataset-construction}

We use the MUBench dataset~\cite{amann2016mubench}, a widely used benchmark for evaluating API misuse detection and fixing. It contains 89 API misuses collected from 33 real-world Java projects and a developer survey. The dataset was constructed through three main sources: (1) filtering existing bug datasets such as BugClassify, Defects4J, iBugs, and QACrashFix for API misuse cases, (2) mining repositories on SourceForge and GitHub for misuses of Java Cryptography Extension APIs, and (3) conducting a developer survey to identify problems caused by Java API misuse.

Most misuse cases are paired with a proper usage example and a YAML metadata file. The metadata outlines key information about the misuse, including its source, the project name, and the commit in which the issue was fixed. However, the dataset exhibits several structural inconsistencies. For example, some misuse cases provide a single correct usage example but are associated with multiple YAML files, each describing a different misuse scenario linked to that same usage. In a few cases, YAML specifications exist without any corresponding proper usage examples. 

Three authors, each with over four years of professional Java development experience, participated in constructing examples of API misuse for our experiment. They carefully examined the correct usage examples and associated YAML metadata in the MUBench dataset to understand the intended and misused API behaviors. From the dataset, we identified 142 distinct correct usage examples, each linked to a single misuse specification in a YAML file.
For each of these, one author manually created a misuse example using the provided correct usage and the YAML-defined violation. Each misuse was then independently reviewed and validated by another author to ensure correctness and alignment with the misuse intent. In rare cases involving ambiguity, the first author, who has over 15 years of Java development experience, was consulted to resolve confusion and confirm the integrity of the example.
Fig. \ref{fig:misuse-example-using-yaml} presents a representative instance, including (a) the correct API usage (Fig. \ref{fig:correct-example}), (b) the misuse specification (Fig. \ref{fig:misuse-yml}), and (c) the resulting misuse example (Fig.~\ref{fig:misuse-example}).

To improve the scale and diversity of our dataset, we generated four AI-assisted misuse variants for each manually created example. The same three authors used ChatGPT-4o, ChatGPT o3-mini-high, and Claude 3.7 to produce these variants through prompt-based interaction. Each prompt included the correct usage, and the \textit{violations} and \textit{description} fields from the corresponding YAML file. All AI-generated examples were manually validated by two authors to ensure semantic correctness and to eliminate exact clones across the five examples (one manual + four AI-generated).

In five cases where a correct usage example was associated with multiple misuse specifications, we manually created one misuse for each specification and did not generate AI variants to prevent excessive variation. In cases where only a misuse specification was available (i.e., no correct usage was provided), we examined the fixing commit (when accessible) to understand the intended usage and created one misuse example based on the metadata. Again, AI-generated variants were omitted due to the absence of a reference usage.

In total, our dataset contains \textbf{147 distinct correct usage examples} and \textbf{740 API misuse instances}, all of which, whether manually or AI-generated, were individually verified to ensure quality, correctness, and uniqueness.

\subsection{Detection of API Misuse Using GitHub Copilot (RQ1)}
\label{subsec:methodology-detection}

We evaluate GitHub Copilot's ability to detect API misuses by integrating it as an extension within the Visual Studio Code IDE, with the GPT-4o model enabled.
All 740 API misuse examples, including both manually created and AI-generated variants, were used in this evaluation. Additionally, the 147 correct usage examples were included to assess Copilot's behavior in non-misuse scenarios.

For each misuse case, the corresponding code snippet was uploaded as a standalone file. We first assessed Copilot's autonomous detection capability by asking a general question in the chat interface: \textit{Is there any API misuse in this code?} 
If Copilot failed to identify a misuse, we issued follow-up prompts or contextual hints based on the YAML description of the misuse scenario to examine whether guided interactions could improve detection accuracy.
Correct usage examples were evaluated using the same procedure. However, for these cases, we did not provide any feedback or clarification, regardless of whether Copilot correctly recognized the absence of misuse or mistakenly flagged correct code as erroneous.

Each evaluation session was conducted in the presence of at least two authors, who independently observed and finally recorded Copilot's responses. For each case, we noted whether the misuse was correctly identified, missed, or incorrectly flagged in a correct usage scenario. These observations were systematically documented for further analysis as follows.

\begin{itemize}
    \item \textit{Autonomous Detection (True Positive)}: Targeted API misuse was correctly identified by Copilot without any hints.
    \item \textit{Hint-Assisted Detection (True Positive)}: Targeted API misuse was correctly identified by Copilot after providing contextual hints.
    \item \textit{Missed Detection (False Negative)}: Targeted API misuse not identified, even after hints were provided.
    \item \textit{Correct Non-Detection (True Negative)}: Correct usage where no misuse was flagged, as expected.
    \item \textit{Incorrect Detection (False Positive)}: Correct usage incorrectly flagged by Copilot as an API misuse.
\end{itemize}

We evaluate Copilot’s performance across all examples (740 misuse + 147 correct usage) using standard metrics -- precision, recall, F1-score, and accuracy. We also focus on the misuse examples (740) to report the detection rate (autonomous + hint-assisted) and miss rate, and provide a focused assessment of Copilot's effectiveness in identifying API misuses.

\begin{table*}[ht]
\centering
\caption{API Misuse Categories with Descriptions}
\rowcolors{2}{gray!15}{white}
\resizebox{7in}{!}{%
\begin{tabular}{cp{4cm}p{14cm}c}
\toprule
\textbf{ID} & \textbf{Category} & \textbf{Description} & \textbf{Count} \\
\textbf{T1} & missing/condition/value\_or\_state & A required condition or specific value/state check is missing. & 187\\
\textbf{T2} & missing/call & A necessary function or method call is omitted, causing incorrect behavior or an incomplete operation. & 172\\
\textbf{T3} & missing/condition/null\_check & A null check is missing before using a potentially null object (may lead to runtime exceptions). & 120\\
\textbf{T4} & missing/exception\_handling & The code lacks exception handling where there could exist possible runtime errors. & 116\\
\textbf{T5} & missing/call AND redundant/call & One required call is missing, while another unnecessary or duplicate call is present. & 59\\
\textbf{T6} & missing/condition/value\_or\_state AND missing/call & Both a condition check and a required function call are absent. & 40\\
\textbf{T7} & redundant/call & An unnecessary or repeated function/method call exists. & 13\\
\textbf{T8} & missing/condition/synchronization & The code lacks synchronization checks needed for safe concurrent access, potentially causing race conditions. & 9\\
\textbf{T9} & missing/condition/null\_check AND redundant/condition/null\_check & The code inconsistently handles null checks or other conditions—missing one where needed and adding another where it is not. & 7\\
\textbf{T10} & redundant/condition/null\_check & A null check is performed unnecessarily, either duplicating a previous check or being irrelevant in the current context. & 6\\
\textbf{T11} & redundant/condition/value\_or\_state & A value/state condition is checked redundantly. & 2\\
\textbf{T12} & redundant/exception\_handling & The code includes unnecessary or excessive exception handling, such as catching exceptions that will not occur. & 2\\
\textbf{T13} & redundant/iteration & The code performs iteration that is unnecessary or repeated (leads to performance issues or logical errors). & 2\\
\textbf{T14} & missing/iteration & An expected loop or iteration over a set of elements is not present. & 1\\
\textbf{T15} & redundant/condition/context & A contextual condition is unnecessarily checked, duplicating checks already guaranteed in the existing code. & 1\\
\textbf{T16} & redundant/condition/synchronization & Synchronization logic is used even when not needed, adding complexity and performance costs. & 1\\
\textbf{T17} & missing/condition/context & A necessary contextual condition (e.g., state or scope) is not checked, which may allow unexpected behavior in an undesired context. & 1\\
\textbf{T18} & redundant/call/duplicate & The same function or method is called multiple times without need, resulting in duplicated execution or unpredictable side effects. & 1\\ \bottomrule
\end{tabular}
}
\label{table:misuse-category-details}
\end{table*}

\subsection{Category-wise Misuse Detection (RQ2)}
\label{subsec:category-wise-detection}

We categorize the API misuse examples based on their violation types as specified in the YAML metadata (e.g., in Fig. \ref{fig:misuse-yml}, the violation type is \textit{missing/condition/null\_check}). Table \ref{table:misuse-category-details} summarizes all violation categories along with their respective counts.
For each category, we analyze Copilot’s detection rate (autonomous + hint-assisted) and miss rate to evaluate how effectively it identifies different types of API misuses. The results are reported per violation category to highlight variations in detection performance across misuse types.


\subsection{Fixing API Misuse Using GitHub Copilot}
\label{subsec:misuse-fixing}

We evaluate GitHub Copilot’s ability to fix API misuses where they were correctly detected (either autonomously or after hints) in the previous experiment. For each case, we provided the misused code and prompted: \textit{Can you correct the API misuse, if any?}
Two authors manually reviewed the fixed code, comparing it against the ground truth to assess whether the misuse was correctly addressed and the original program semantics were preserved. We excluded all cases where the misuse was not detected or detected incorrectly, as Copilot would have no basis for initiating a fix. We also excluded correct usage examples, as fixing where no misuse exists is neither meaningful nor expected.
Each case was documented as either correctly fixed or incorrectly fixed, and we report the overall fixing rate along with category-wise breakdowns based on violation types.

\section{Study Findings}
\label{sec:study-findings}

We present key findings on GitHub Copilot’s ability to detect and fix API misuses. We first report overall detection performance using standard metrics, followed by a category-wise analysis. We then assess Copilot’s effectiveness in fixing misuses it successfully identified.

\subsection{Findings on Overall Detection Performance (RQ1)}
\label{subsec:findings-rq1}

\begin{figure}
	\centering
	\resizebox{2in}{!}{
    \begin{tikzpicture}
    \pie[explode=0.3, text=pin, number in legend, sum = auto, color={black!10, black!50, black!90}]
        { 86.75/\Huge{AD (86.75\%)},
          5.68/\Huge{HAD (5.68\%)},
          7.57/\Huge{MD (7.57\%)}
          }
    \end{tikzpicture}
    }
	\caption{Findings summary of API misuse detection (\textbf{AD:} \underline{A}utonomous \underline{D}etection, \textbf{HAD:} \underline{H}int-\underline{A}ssisted \underline{D}etection, \textbf{MD:} \underline{M}issed \underline{D}etection).}
	\label{fig:piechart-api-misuse-summary}
	\vspace{-5mm}
\end{figure}
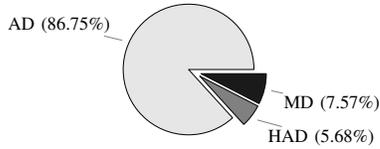

Out of 887 total examples (740 misuse + 147 correct usage), GitHub Copilot demonstrates a promising ability to identify API misuses, even without guidance. 
Fig.~\ref{fig:piechart-api-misuse-summary} summarizes the overall performance of GitHub Copilot in detecting API misuses across 740 misuse examples. A large majority of cases, 642 (86.75\%), were successfully detected without any hint (Autonomous Detection), and 42 (5.68\%) required contextual guidance (Hint-Assisted Detection). However, 56 misuses went undetected despite hints. Sixty-six correct usage examples were incorrectly flagged as misuses. Interestingly, 81 correct usage examples were left unflagged, as they should be, indicating Copilot’s capacity to avoid over-correction in many cases.
This result suggests that while Copilot is highly effective at detecting certain API misuses on its own, it also exhibits over-sensitivity in some correct usage contexts and fails to recognize more nuanced misuse patterns without assistance.

\begin{table}[ht]
\centering
\caption{Confusion Matrix of API Misuse Detection}
\resizebox{2.8in}{!}{
\begin{tabular}{l|c|c}
\multicolumn{1}{c|}{} & \textbf{Predicted Misuse} & \begin{tabular}[c]{@{}c@{}}Predicted Correct / \\Missed Targeted Misuse\end{tabular} \\
\toprule
\textbf{Actual Misuse} & 684 (TP) & 56 (FN) \\
\textbf{Actual Correct} & 66 (FP) & 81 (TN) \\
\midrule
\end{tabular}
}
\label{tab:confusion-matrix}
\end{table}



Based on the values in Table~\ref{tab:confusion-matrix}, we compute the standard performance metrics as follows: Precision = 91.2\%, Recall = 92.4\%, F1-Score = 91.8\%, and Overall Accuracy = 86.2\%.


\begin{findingbox}
\leftskip 8pt \rightskip 2pt \textbf{Summary of RQ\textsubscript{1}}: 
Copilot detected API misuses with 91.2\% precision and 91.8\% F1-score, showing strong potential for identifying critical issues. However, its 45\% false positive rate on correct code highlights the need for better context awareness.
\end{findingbox}

\begin{figure}[htbp]
\centering
\resizebox{3.2in}{!}{
\begin{tikzpicture}
\begin{axis}[
    xbar stacked,
    bar width=10pt,
    width=5in,
    height=4in,
    ymin=0.5, ymax=18.8,
    ytick={1,...,18},
    yticklabels={
        T18,
        T17,
        T16,
        T15,
        T14,
        T13,
        T12,
        T11,
        T10,
        T9,
        T8,
        T7,
        T6,
        T5,
        T4,
        T3,
        T2,
        T1
    },
    xtick={0,20,...,120},
    xticklabels={0\%,20\%,40\%,60\%,80\%,100\%,120\%},
    enlarge y limits=0.01,
    nodes near coords,
    point meta=explicit,
    visualization depends on={value \thisrow{meta} \as \perval},
    every node near coord/.append style={
        font=\scriptsize,
        yshift=0pt,
        text=black,
    },
    nodes near coords style={
        /pgf/number format/fixed,
        /pgf/number format/precision=1
    },
    nodes near coords={%
        \ifdim\perval pt>4.99pt
            \pgfmathprintnumber{\perval}\%%
        \else
        \fi
    },
    tick label style={font=\large},
    y label style={font=\small},
    label style={font=\small},
    legend style={
        at={(0.5,-0.06)},
        anchor=north,
        legend columns=3,
        font=\small,
        draw=none
    }
]

\addplot+[xbar, fill=green!70] table [meta=meta, row sep=\\] {
x y meta \\
100 1 100 \\
100 2 100 \\
100 3 100 \\
0   4 0 \\
0   5 0 \\
100 6 100 \\
100 7 100 \\
0   8 0 \\
83.33 9 83.33 \\
100 10 100 \\
77.78 11 77.78 \\
53.85 12 53.85 \\
95 13 95 \\
93.22 14 93.22 \\
87.07 15 87.07 \\
93.33 16 93.33 \\
88.37 17 88.37 \\
80.75 18 80.75 \\
};

\addplot+[xbar, fill=orange!80] table [meta=meta, row sep=\\] {
x y meta \\
0 1 0 \\
0 2 0 \\
0 3 0 \\
100 4 100 \\
0 5 100 \\
0 6 0 \\
0 7 0 \\
50 8 50 \\
0 9 0 \\
0 10 0 \\
0 11 0 \\
23.08 12 23.08 \\
0 13 0 \\
1.69 14 1.69 \\
7.76 15 7.76 \\
1.67 16 1.67 \\
4.65 17 4.65 \\
9.09 18 9.09 \\
};

\addplot+[xbar, fill=red!70] table [meta=meta, row sep=\\] {
x y meta \\
0 1 0 \\
0 2 0 \\
0 3 0 \\
0 4 0 \\
100 5 100 \\
0 6 0 \\
0 7 0 \\
50 8 50 \\
16.67 9 16.67 \\
0 10 0 \\
22.22 11 22.22 \\
23.08 12 23.08 \\
5 13 5 \\
5.08 14 5.08 \\
5.17 15 5.17 \\
5 16 5 \\
6.98 17 6.98 \\
10.16 18 10.16 \\
};

\node[anchor=west, font=\footnotesize] at (102,1) {\textbf{1}};
\node[anchor=west, font=\footnotesize] at (102,2) {\textbf{1}};
\node[anchor=west, font=\footnotesize] at (102,3) {\textbf{1}};
\node[anchor=west, font=\footnotesize] at (102,4) {\textbf{1}};
\node[anchor=west, font=\footnotesize] at (102,5) {\textbf{1}};
\node[anchor=west, font=\footnotesize] at (102,6) {\textbf{2}};
\node[anchor=west, font=\footnotesize] at (102,7) {\textbf{2}};
\node[anchor=west, font=\footnotesize] at (102,8) {\textbf{2}};
\node[anchor=west, font=\footnotesize] at (102,9) {\textbf{6}};
\node[anchor=west, font=\footnotesize] at (102,10) {\textbf{7}};
\node[anchor=west, font=\footnotesize] at (102,11) {\textbf{9}};
\node[anchor=west, font=\footnotesize] at (102,12) {\textbf{13}};
\node[anchor=west, font=\footnotesize] at (102,13) {\textbf{40}};
\node[anchor=west, font=\footnotesize] at (102,14) {\textbf{59}};
\node[anchor=west, font=\footnotesize] at (102,15) {\textbf{116}};
\node[anchor=west, font=\footnotesize] at (102,16) {\textbf{120}};
\node[anchor=west, font=\footnotesize] at (102,17) {\textbf{172}};
\node[anchor=west, font=\footnotesize] at (102,18) {\textbf{187}};

\legend{Autonomous Detection, Hint-Assisted Detection, Missed Detection}

\end{axis}
\end{tikzpicture}
}
\caption{Detection outcomes for different API misuse types (only values $>=$5\% labeled) (see Table \ref{table:misuse-category-details} for T1--T18 details).}
\label{fig:category-wise-api-misuse-detection}
\end{figure}
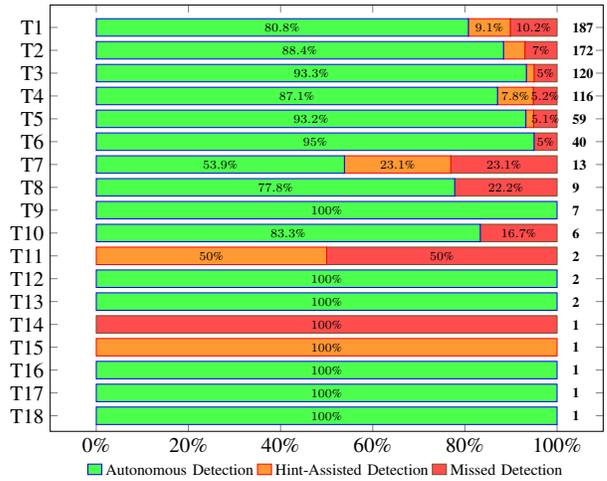

\subsection{Findings on Category-wise Detection Analysis (RQ2)}
\label{subsec:findings-rq2}

To gain deeper insight, we analyze detection outcomes across individual misuse categories (Fig.~\ref{fig:category-wise-api-misuse-detection}). Detection rates vary notably by violation type:

\begin{itemize}
    \item Common misuse types such as missing/call, missing/condition/null\_check, and missing/exception\_handling exhibit high autonomous detection rates, exceeding 85\%. These patterns are likely easier for Copilot to identify, possibly due to their frequent occurrence in training data and clear syntactic cues.

    \item In contrast, more nuanced or semantically dependent violations, such as redundant/call, missing/condition/value\_or\_state AND missing/call, or missing/condition/synchronization, show lower detection rates or require hints. These misuses often involve implicit logic, cross-method dependencies, or non-obvious behavioral contracts, making them harder to flag without a deeper understanding of the broader context.

    \item Similarly, synchronization-related errors demand awareness of concurrency assumptions, thread safety guarantees, or life cycle constraints, areas where current LLM-based models like Copilot may have limited semantic modeling capabilities.
\end{itemize}

Fig.~\ref{fig:category-wise-api-misuse-detection} reveals this disparity, indicating that Copilot performs reliably on common, structurally explicit misuses, but struggles with compound, context-sensitive, or less frequently represented misuse types. These findings suggest that while Copilot is a powerful assistant for routine misuse detection, its reasoning remains local and mainly syntactic, limiting its ability to catch more complex or subtle violations.

\begin{findingbox}
\leftskip 8pt \rightskip 2pt \textbf{Summary of RQ\textsubscript{2}}: 
Out of 740 API misuses, Copilot detected 86.75\% autonomously and 5.68\% with hints, missing 7.57\% even after prompting. While it excelled at common patterns like missing/call and missing/null\_check, its performance dropped on complex or context-sensitive misuses, revealing limitations in deeper semantic reasoning.
\end{findingbox}

\subsection{Findings on Fixing Effectiveness for Detected Misuses (RQ3)}
\label{subsec:findings-rq3}


To evaluate GitHub Copilot's ability to not only detect but also correct API misuses, we analyzed its performance on cases where the misuse was correctly identified. The goal was to assess whether Copilot could generate semantically correct fixes that align with the intended program behavior.
Our findings show a high fixing success rate for common misuse categories. Specifically, Copilot achieved 98.8\% accuracy in fixing missing/call (168 cases), 99.1\% for missing/condition/null\_check (160 cases), 87.3\% for missing/exception\_handling (114 cases), and 94.6\% for missing/condition/value\_or\_state (168 cases). These results suggest that Copilot can effectively fix misuse cases that follow frequent or well-documented patterns, likely due to the presence of consistent syntactic cues in the training data.

Additionally, Copilot attained 100\% fixing accuracy in several low-frequency categories, such as redundant/call, missing/condition/synchronization, and redundant/condition/null\_check, though these categories had small sample sizes ranging from 1 to 10 instances. While encouraging, the generalizability of these perfect scores should be interpreted with caution due to the limited data.
However, Copilot's performance dropped slightly for more complex or compound misuse patterns. For example, it achieved only 85.7\% accuracy on missing/condition/null\_check AND redundant/condition/null\_check (7 cases), 90\% on missing/condition/value\_or\_state AND missing/call (38 cases), and just 50\% on redundant/iteration (2 cases). These patterns typically involve subtle program logic, deeper semantic understanding, or implicit API usage constraints that go beyond surface-level patterns.

\begin{findingbox}
\leftskip 8pt \rightskip 2pt \textbf{Summary of RQ\textsubscript{3}}:
Copilot correctly fixed over \textbf{95\%} of the API misuses it identified, with particularly high accuracy for common issues like \textit{missing/call} and \textit{null\_check}. This highlights its strong ability to fix well-known patterns, but it still struggles with cases requiring broader contextual reasoning or deeper semantic interpretation.
\end{findingbox}

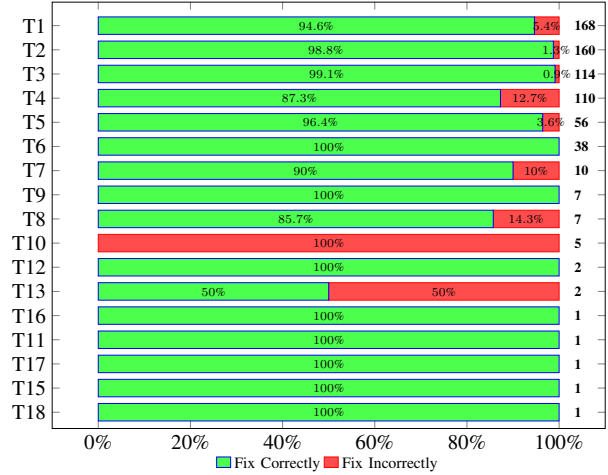
\begin{figure}[htbp]
\centering
\resizebox{3.2in}{!}{
\begin{tikzpicture}
\begin{axis}[
    xbar stacked,
    bar width=10pt,
    width=5in,
    height=4in,
    ymin=0.5, ymax=17.8,
    ytick={1,...,17},
    yticklabels={
        T18,
        T15,
        T17,
        T11,
        T16,
        T13,
        T12,
        T10,
        T8,
        T9,
        T7,
        T6,
        T5,
        T4,
        T3,
        T2,
        T1
    },
    xtick={0,20,...,120},
    xticklabels={0\%,20\%,40\%,60\%,80\%,100\%,120\%},
    enlarge y limits=0.01,
    nodes near coords,
    point meta=explicit,
    visualization depends on={value \thisrow{meta} \as \perval},
    every node near coord/.append style={
        font=\scriptsize,
        yshift=0pt,
        text=black,
    },
    nodes near coords style={
        /pgf/number format/fixed,
        /pgf/number format/precision=1
    },
    nodes near coords =\pgfmathprintnumber{\pgfplotspointmeta}\%,
    tick label style={font=\large},
    y label style={font=\small},
    label style={font=\small},
    legend style={
        at={(0.5,-0.05)},
        anchor=north,
        legend columns=2,
        font=\small,
        draw=none
    }
]

\addplot+[xbar, fill=green!70] table [meta=meta, row sep=\\] {
x y meta \\
100 1 100 \\
100 2 100 \\
100 3 100 \\
100   4 100 \\
100   5 100 \\
50 6 50 \\
100 7 100 \\
0   8 0 \\
85.71 9 85.71 \\
100 10 100 \\
90 11 90 \\
100 12 100 \\
96.43 13 96.43 \\
87.27 14 87.27 \\
99.12 15 99.12 \\
98.75 16 98.75 \\
94.64 17 94.64 \\
};

\addplot+[xbar, fill=red!70] table [meta=meta, row sep=\\] {
x y meta \\
0 1 0 \\
0 2 0 \\
0 3 0 \\
0 4 0 \\
0 5 0 \\
50 6 50 \\
0 7 0 \\
100 8 100 \\
14.29 9 14.29 \\
0 10 0 \\
10 11 10 \\
0 12 0 \\
3.57 13 3.57 \\
12.73 14 12.73 \\
0.88 15 0.88 \\
1.25 16 1.25 \\
5.36 17 5.36 \\
};

\node[anchor=west, font=\footnotesize] at (102,1) {\textbf{1}};
\node[anchor=west, font=\footnotesize] at (102,2) {\textbf{1}};
\node[anchor=west, font=\footnotesize] at (102,3) {\textbf{1}};
\node[anchor=west, font=\footnotesize] at (102,4) {\textbf{1}};
\node[anchor=west, font=\footnotesize] at (102,5) {\textbf{1}};
\node[anchor=west, font=\footnotesize] at (102,6) {\textbf{2}};
\node[anchor=west, font=\footnotesize] at (102,7) {\textbf{2}};
\node[anchor=west, font=\footnotesize] at (102,8) {\textbf{5}};
\node[anchor=west, font=\footnotesize] at (102,9) {\textbf{7}};
\node[anchor=west, font=\footnotesize] at (102,10) {\textbf{7}};
\node[anchor=west, font=\footnotesize] at (102,11) {\textbf{10}};
\node[anchor=west, font=\footnotesize] at (102,12) {\textbf{38}};
\node[anchor=west, font=\footnotesize] at (102,13) {\textbf{56}};
\node[anchor=west, font=\footnotesize] at (102,14) {\textbf{110}};
\node[anchor=west, font=\footnotesize] at (102,15) {\textbf{114}};
\node[anchor=west, font=\footnotesize] at (102,16) {\textbf{160}};
\node[anchor=west, font=\footnotesize] at (102,17) {\textbf{168}};

\legend{Fix Correctly, Fix Incorrectly}

\end{axis}
\end{tikzpicture}
}
\caption{Fixing summaries for different API misuse types (see Table \ref{table:misuse-category-details} for T1--T18 details).}
\label{fig:api-misuse-fixing}
\end{figure}

\section{Related Work}
\label{sec:related-work}

Prior studies have extensively evaluated GitHub Copilot's capabilities, primarily focusing on program synthesis \cite{mastropaolo2023robustness, moassessing} and assessing the quality of its generated code \cite{yetistiren2022assessing, imai2022github, nguyen2022empirical, dakhel2023github, wermelinger2023using, moassessing}. Additionally, a few studies explored Copilot's effectiveness in generating test cases \cite{mehmood2023manual, el2024using, alves2024detecting}. While these studies highlight Copilot's evident strengths in program synthesis, pair programming, and test case generation, its capabilities in other critical software engineering areas remain largely unexplored. Notably, the use of Copilot for API misuse detection and correction in real-world contexts has not yet been systematically investigated.

API misuse remains a major concern, as developers often rely on unreliable sources like Stack Overflow or AI-generated code, both shown to contain high rates of misuse (31\% and 70\%, respectively) \cite{10.1145/3180155.3180260, mousavi2024investigationmisusejavasecurity}. In response, various static and dynamic tools (e.g., MisuseHint, ASAP-Repair, Hotfixer) have been developed to detect and repair such issues. However, benchmark studies like \textit{CamBench} reveal persistent problems, including high false positives and limited contextual awareness.
While LLMs such as ChatGPT offer promise in identifying subtle misuses, they still struggle with hallucinations and lack transparency. Most tools, including AI-based ones, function post-development and are not integrated into the coding workflow, which limits their ability to provide timely and actionable feedback.

Our study addresses this gap by evaluating GitHub Copilot’s ability to detect and fix real-time API misuses directly within the IDE. To our knowledge, this is the first empirical analysis using the MUBench benchmark to detect and fix API misuse. The findings demonstrate Copilot’s potential as an inline, misuse-aware coding assistant that supports developers during code authoring.

\section{Threats to Validity}
\label{sec:threats-to-validity}



Threats to \textbf{external validity} concern the generalizability of our findings. We used the MUBench dataset, which includes diverse Java API misuses. While this diversity may support generalization to other languages, the dataset’s age raises the possibility that its content has been seen by LLMs (e.g., GPT models). As a result, our high detection and correction performance may not hold on newer or unseen benchmarks. We encourage caution in generalizing these findings and recommend replication with more recent and varied datasets.

Threats to \textbf{internal validity} relate to potential biases in data preparation and analysis. Our misuse examples were created manually and expanded using AI tools, which introduces risks of error or inconsistency. To mitigate this, all examples were independently verified by at least two authors, with disagreements resolved by a highly experienced reviewer. We also followed a consistent protocol for evaluating Copilot’s responses, reducing the likelihood of procedural bias.

Threats to \textbf{construct validity} involve the suitability of our evaluation approach. We used widely accepted metrics (precision, recall, F1-score, accuracy) to assess Copilot’s performance. However, some misuse categories had limited examples, which may impact the strength of our findings of category-specific conclusions.

\section{Conclusion}
\label{sec:conclusion}

We evaluated GitHub Copilot’s effectiveness in detecting and fixing API misuse using the MUBench dataset. Copilot achieved high detection accuracy (86.2\%), precision (91.2\%), and recall (92.4\%), and successfully corrected over 95\% of the misuses it identified—particularly for common patterns like missing method calls and null checks.
However, Copilot struggled with complex or context-sensitive cases requiring deeper semantic understanding. Despite this, it shows strong potential as a real-time assistant that complements traditional analysis tools by providing immediate, in-context feedback.
Future work should explore broader datasets, additional languages, and hybrid techniques to enhance Copilot’s reasoning and generalizability.

\section*{Acknowledgment}
This research is supported in part by the Natural Sciences and Engineering Research Council of Canada (NSERC) Discovery Grants program, the Canada Foundation for Innovation's John R. Evans Leaders Fund (CFI-JELF), and by the industry-stream NSERC CREATE in Software Analytics Research (SOAR).

\balance

\bibliographystyle{unsrtnat}
\footnotesize
\bibliography{bibliography}

\begin{thebibliography}{22}
\providecommand{\natexlab}[1]{#1}
\providecommand{\url}[1]{\texttt{#1}}
\expandafter\ifx\csname urlstyle\endcsname\relax
  \providecommand{\doi}[1]{doi: #1}\else
  \providecommand{\doi}{doi: \begingroup \urlstyle{rm}\Url}\fi

\bibitem[Sven et~al.(2019)Sven, Nguyen, Nadi, Nguyen, and Mezini]{sven2019investigating}
A.~Sven, H.~A. Nguyen, S.~Nadi, T.~N. Nguyen, and M.~Mezini.
\newblock Investigating next steps in static api-misuse detection.
\newblock In \emph{In Proc. MSR}, pages 265--275, 2019.

\bibitem[Wei et~al.(2024)Wei, Harzevili, Huang, Yang, Wang, and Wang]{wei2024demystifying}
M.~Wei, N.~S. Harzevili, Y.~Huang, J.~Yang, J.~Wang, and S.~Wang.
\newblock Demystifying and detecting misuses of deep learning apis.
\newblock In \emph{In Proc. ICSE}, pages 1--12, 2024.

\bibitem[Li et~al.(2021)Li, Jiang, Benton, Xiong, and Zhang]{li2021large}
X.~Li, J.~Jiang, S.~Benton, Y.~Xiong, and L.~Zhang.
\newblock A large-scale study on api misuses in the wild.
\newblock In \emph{In Proc. ICST}, pages 241--252, 2021.

\bibitem[Krasner(2022)]{Krasner_2022}
H.~Krasner.
\newblock The cost of poor software quality in the us: A 2022 report, 2022.
\newblock URL \url{https://www.it-cisq.org/wp-content/uploads/sites/6/2022/11/CPSQ-Report-Nov-22-2.pdf}.

\bibitem[Mastropaolo et~al.(2023)Mastropaolo, Pascarella, Guglielmi, Ciniselli, Scalabrino, Oliveto, and Bavota]{mastropaolo2023robustness}
A.~Mastropaolo, L.~Pascarella, E.~Guglielmi, M.~Ciniselli, S.~Scalabrino, R.~Oliveto, and G.~Bavota.
\newblock On the robustness of code generation techniques: An empirical study on github copilot.
\newblock In \emph{In Proc. ICSE}, pages 2149--2160, 2023.

\bibitem[Mo et~al.(2025)Mo, Wang, Zhan, Jiang, Wang, Zhao, Li, and Ma]{moassessing}
R.~Mo, D.~Wang, W.~Zhan, Y.~Jiang, Y.~Wang, Y.~Zhao, Z.~Li, and Y.~Ma.
\newblock Assessing and analyzing the correctness of github copilot’s code suggestions.
\newblock \emph{TOSEM}, 2025.

\bibitem[Yetistiren et~al.(2022)Yetistiren, Ozsoy, and Tuzun]{yetistiren2022assessing}
B.~Yetistiren, I.~Ozsoy, and E.~Tuzun.
\newblock Assessing the quality of github copilot’s code generation.
\newblock In \emph{In Proc. PROMISE}, pages 62--71, 2022.

\bibitem[Imai(2022)]{imai2022github}
S.~Imai.
\newblock Is github copilot a substitute for human pair-programming? an empirical study.
\newblock In \emph{In Proc. ICSE-Companion}, pages 319--321, 2022.

\bibitem[Nguyen and Nadi(2022)]{nguyen2022empirical}
N.~Nguyen and S.~Nadi.
\newblock An empirical evaluation of github copilot's code suggestions.
\newblock In \emph{In Proc. MSR}, pages 1--5, 2022.

\bibitem[Dakhel et~al.(2023)Dakhel, Majdinasab, Nikanjam, Khomh, Desmarais, and Jiang]{dakhel2023github}
A.~M. Dakhel, V.~Majdinasab, A.~Nikanjam, F.~Khomh, M.~C. Desmarais, and Z.~M.~J. Jiang.
\newblock Github copilot ai pair programmer: Asset or liability?
\newblock \emph{JSS}, 203, 2023.

\bibitem[Wermelinger(2023)]{wermelinger2023using}
M.~Wermelinger.
\newblock Using github copilot to solve simple programming problems.
\newblock In \emph{In Proc. SIGCSE V.1}, pages 172--178, 2023.

\bibitem[Mehmood et~al.(2023)Mehmood, Janjua, and Ahmed]{mehmood2023manual}
S.~Mehmood, U.~I. Janjua, and A.~Ahmed.
\newblock From manual to automatic: The evolution of test case generation methods and the role of github copilot.
\newblock In \emph{In Proc. FIT}, pages 13--18, 2023.

\bibitem[E~Haji et~al.(2024)E~Haji, Brandt, and Zaidman]{el2024using}
K.~E~Haji, C.~Brandt, and A.~Zaidman.
\newblock Using github copilot for test generation in pythoncopilot’s: An empirical study.
\newblock In \emph{In Proc. AST}, pages 45--55, 2024.

\bibitem[Alves et~al.(2024)Alves, Santos, Bezerra, and Machado]{alves2024detecting}
V.~A. Alves, C.~Santos, C.~Bezerra, and I.~Machado.
\newblock Detecting test smells in python test code generated by llm: An empirical study with github copilot.
\newblock In \emph{In Proc. SBES}, pages 581--587, 2024.

\bibitem[Liang et~al.(2022)Liang, Kuai, Zhang, Zhang, Kuang, and Zhang]{9885793}
Q.~Liang, Z.~Kuai, Y.~Zhang, Z.~Zhang, L.~Kuang, and L.~Zhang.
\newblock Misusehint: A service for api misuse detection based on building knowledge graph from documentation and codebase.
\newblock In \emph{In Proc. ICWS}, pages 246--255, 2022.

\bibitem[Masood and Martin(2024)]{masood2024statictoolsevaluatinglarge}
Zohaib Masood and Miguel~Vargas Martin.
\newblock Beyond static tools: Evaluating large language models for cryptographic misuse detection, 2024.
\newblock URL \url{https://arxiv.org/abs/2411.09772}.

\bibitem[Xia et~al.(2024)Xia, Xie, Liu, Lu, Liu, Wang, and Ji]{xia2024exploringautomaticcryptographicapi}
Yifan Xia, Zichen Xie, Peiyu Liu, Kangjie Lu, Yan Liu, Wenhai Wang, and Shouling Ji.
\newblock Exploring automatic cryptographic api misuse detection in the era of llms, 2024.
\newblock URL \url{https://arxiv.org/abs/2407.16576}.

\bibitem[Dawson et~al.(2010)Dawson, Burrell, Rahim, and Brewster]{Dawson2010}
Maurice Dawson, Darrell Burrell, Emad Rahim, and Stephen Brewster.
\newblock Integrating software assurance into the software development life cycle (sdlc).
\newblock \emph{Journal of Information Systems Technology and Planning}, 3:\penalty0 49--53, 01 2010.

\bibitem[Amann et~al.(2016)Amann, Kämmerer, Nadi, Nguyen, Nguyen, and Schlitzer]{amann2016mubench}
Sven Amann, Mattis Kämmerer, Sarah Nadi, Hoan~Anh Nguyen, Tien~N. Nguyen, and Jonas Schlitzer.
\newblock Mubench: A benchmark for api-misuse detectors, 2016.
\newblock URL \url{https://github.com/stg-tud/MUBench}.
\newblock Accessed: 2025-03-07.

\bibitem[Mondal et~al.(2025)Mondal, Roy, Wang, Arguello, and Mathan]{replication}
Saikat Mondal, Chanchal~K. Roy, Hong Wang, Juan Arguello, and Samantha Mathan.
\newblock Replication package, 2025.
\newblock URL \url{https://figshare.com/s/074ddd0b15a0148107b0}.

\bibitem[Zhang et~al.(2018)Zhang, Upadhyaya, Reinhardt, Rajan, and Kim]{10.1145/3180155.3180260}
T.~Zhang, G.~Upadhyaya, A.~Reinhardt, H.~Rajan, and M.~Kim.
\newblock Are code examples on an online q\&a forum reliable? a study of api misuse on stack overflow.
\newblock In \emph{Proc. ICSE}, page 886–896, 2018.

\bibitem[Mousavi et~al.(2024)Mousavi, Islam, Moore, Abuadbba, and Babar]{mousavi2024investigationmisusejavasecurity}
Zahra Mousavi, Chadni Islam, Kristen Moore, Alsharif Abuadbba, and Muhammad~Ali Babar.
\newblock An investigation into misuse of java security apis by large language models, 2024.
\newblock URL \url{https://arxiv.org/abs/2404.03823}.

\end{thebibliography}

\end{document}